\documentclass[a4paper]{jpconf}
\usepackage{graphicx}
\begin{document}
\title{Abnormal magnetoresistance behavior in Nb thin film with rectangular antidot lattice}

\author{W J Zhang, S K He, B H Li, F Cheng, B Xu, Z C Wen, W H Cao, X F Han, S P Zhao, X G Qiu}

\address{Beijing National Laboratory for Condensed Matter Physics, Institute of Physics, Chinese Academy of Sciences, P.O. Box 603, Beijing 100190, China}

\ead{zhangweijun@ssc.iphy.ac.cn, xgqiu@iphy.ac.cn}

\begin{abstract}
Abnormal magnetoresistance behavior is found in superconducting Nb films perforated with rectangular arrays of antidots (holes). Generally magnetoresistance were always found to increase with increasing magnetic field. Here we observed a reversal of this behavior for particular in low temperature or current density. This phenomenon is due to a strong 'caging effect' which interstitial vortices are strongly trapped among pinned multivortices.
\end{abstract}

\section{Introduction}
Superconducting thin films with periodic arrays of pinning sites have received much attention in purposes to enhance critical parameters\cite{Harada_sci_96,Moshchalkov_prb_1998,Hoffmann_prb_2000,Silhanek_prb_2005} and artificially control the motion of vortices\cite{RevModPhys.81.387}. When the number of vortices equals to an integer multiple or fractional of the number of pinning centers, dips in resistance or peaks in critical current as a function of the applied magnetic field can be visible, which was known as the commensurate effect or matching effect\cite{Harada_sci_96,Field_prl_2002}.
In Nb thin films with rectangular array of magnetic dots and nonmagnetic dots, interesting phenomena have been revealed, such as the channeling effect\cite{Villegas_prb_2005}, anisotropy in critical current\cite{Reichhardt_prb_2001} and vortex-lattice reconfiguration transition\cite{Artificially,Temperature,Hysteresis}. When the reconfiguration transition occurred, changes in the shape of the minima and their periodicity in the magnetoresistance curves were found. Two possible models have been proposed to explain this phenomenon: the geometrical reconfiguration model\cite{Artificially}, and the multivortex model\cite{Hysteresis}. However, it's still unclear that the causes of the oscillations in magnetoresistance after the transition. In this paper, to investigate the pinning mechanism in high field regime, we performed transport measurements in rectangular arrays of antidots with different aspect ratios of unit cell. We observed a temperature and current dependent reconfiguration transition. After the transition, a decrease in magnetoresistance was found at high field. This abnormal behavior is accounted for a reduction of the mobility of interstitial vortices by a 'caging effect'. This effect has been predicted by theoretical simulations\cite{Berdiyorov_EPL_2006, Berdiyorov_prb_2006}.
\section{Experiment}
High quality Nb thin films were deposited on Si substrate by magnetron sputtering, with a thickness of about $100$~nm, critical temperature $T_c=8.870$~K and the transition width of about $50$~mK. For transport measurements, two four-probe microbridges were fabricated on one chip with ultraviolet photolithography and then etched by enhanced reactive ion etching. At each center of the bridges, there is a $60$~$\mu$m $\times$ $60$~$\mu$m square area for Electron Beam Lithography writing. The distance between two voltage connects is $60$~$\mu$m.

\begin{table}
\caption{\label{table} Sample characteristics. $\Delta W$ is the nearest separation between two antidots along the short side of rectangular array; $D$ is antidot diameter. $\Delta T_c$ is the resistance transition width.}
\begin{center}
\begin{tabular}{ccccccc}
   \br
   Pattern & $T_c$~(K) & $\Delta T_c$~(K) & $R_N~(\Omega)$ & $\Delta W$~(nm) & $D$~(nm) & $b/a$\\
   \mr
   A (600 $\times$ 1000) & 8.810 & 0.095 & 2.517 & 124 & 476 & 1.67 \\

   B (600 $\times$ 1400) & 8.818 & 0.109 & 2.044 & 121 & 479 & 2.33\\
   \br
 \end{tabular}
 \end{center}
\end{table}
 \begin{figure}
  \begin{center}
    \includegraphics[width=14.5cm]{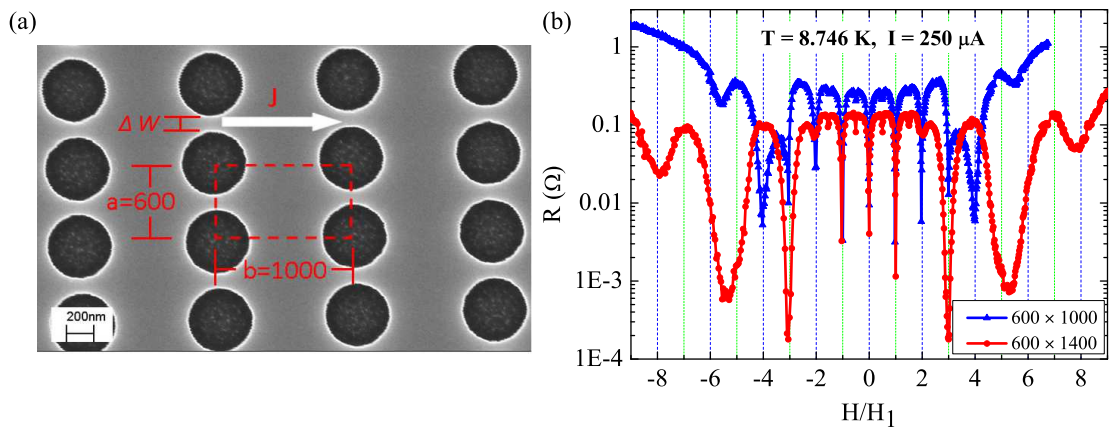}%
    \caption{\label{SEM} (color online) (a) SEM image of a rectangular antidot lattice with an unit cell of $600$~nm $\times$ $1000$~nm. And the hole diameter is $476$~nm. The dash rectangle indicates a unit cell. (b) Field dependence of resistance at $T=8.746$~K and $I=250~\mu$A for pattern A $(600\times 1000)$ and B $(600\times 1400)$. The curves are normalized by the first matching field $H_{1A}=34.0$~Oe or $H_{1B}=24.8$~Oe, respectively.}
  \end{center}
\end{figure}

The measurements were performed in Quantum Design Physical Properties Measurement System (PPMS-14), with the applied magnetic field oriented perpendicular to the film surface. The temperature stability was 2 mK during the measurements. The superconducting coherence length $\xi(0)$ was 10.8 nm and the penetration depth $\lambda(0)$ was 75.1 nm, determined from an unpatterned Nb microbridge by measured its phase boundary $T_c(H)$\cite{Tinkham_book}.

\section{Discussion}
Rectangular arrays of antidots with different aspect ratios of unit cells $a$ $\times$ $b$ were separately patterned on two Nb microbridges. The relevant parameters of the two samples are summarized in Table~\ref{table}. In Figure~\ref{SEM}(a), image of Scanning electron microscope (SEM) shows that the overall periodicity of antidot lattices is maintained very well. Currents flow along side $b$ of the rectangular unit cell. Figure~\ref{SEM}(b) shows the magnetoresitance $R(H)$ curves of pattern A and B. They were both recorded at $T$ = 8.746 K with a current $I$ = 250 $\mu$A. The magnetic field swept from $-$ 300 Oe to $+$ 220 Oe with a step of 0.4 Oe. For both samples, two different regimes can be clearly distinguished from the curves. In low field regime, fractional matching minima (1/3, 1/2 and 2/3)$H_{1}$ and sharp integer matching dips can be observed. The interval $\Delta H_{A}$ = 34.0 $\pm$ 1.4 Oe ($\Delta H_{B}$ = 24.8 $\pm$ 1.0 Oe) between neighbor dips is in good agreement with the theoretical value $\Delta H_{AT}$ = $\Phi_0 /ab$ = 34.5 Oe ($\Delta H_{BT}$ = 24.6 Oe)\cite{Tinkham_book}, where $\Phi_0$ is the flux quantum. It implies that the vortex lattice is commensurate with the rectangular array of antidots. In high field regime, broad dips ($5.4H_{1A}$, $5.1H_{1B}$ and $7.5H_{1B}$) corresponding to the reconfiguration of interstitial vortices to square lattice\cite{Artificially} were found. The interval in this regime was $\Delta H_{highA}$ = 52.6 $\pm$ 2.6 Oe ($\Delta H_{highB}$ = 57.3 $\pm$ 2.7 Oe). It's closed to the theoretical period for square vortex lattice $\Phi_0 /a^2$ = 57.5 Oe, where $a$ = 600 nm.

In the multivortex model, the transition between the low field and high field regime, indicated the formation of interstitial vortices.\cite{Hysteresis} Using an expression $N_S=D/4\xi (T)$\cite{Mkrtchyan.jetp}, we could roughly estimate the saturation number for pattern A and B at $T$ = 8.746 K, which gives $N_{SA}$ $\sim$ 1 and $N_{SB}$ $\sim$ 1. For a periodic array of antidots, $N_{S}$ can be higher due to the vortex-vortex interactions.\cite{Berdiyorov_prb_2006} In our samples, the dense array (pattern A) had a larger $N_{S}$ than sparse one (pattern B), where $N_{SA}$ = 3 and $N_{SB}$ = 2, as shown in Figure \ref{SEM}(b). This is the main difference between the two samples. Therefore, the saturation number not only depends on the hole size but also the geometry of the pinning array. 

\begin{figure}
  \begin{center}
   \includegraphics[width=14.5cm]{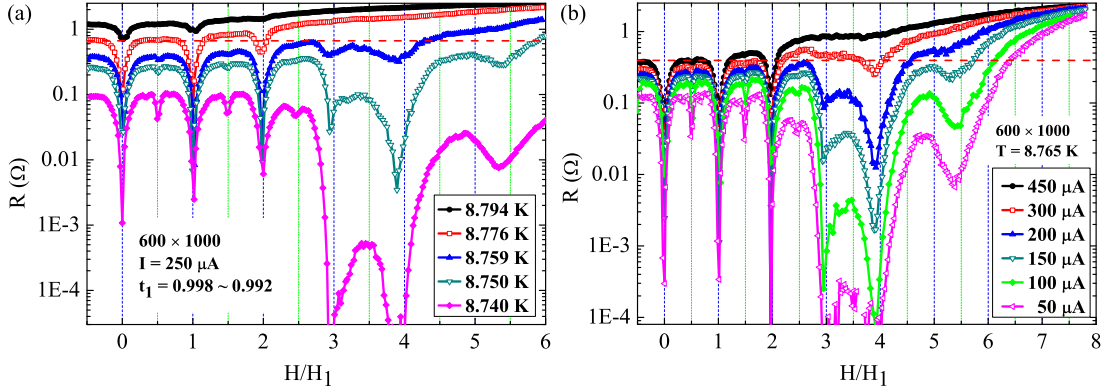}%
   \caption{\label{RH} (color online)
   Normalized magnetoresistance curves of pattern A. (a) With a fixed current $I=250~\mu$A, at several temperatures. (b) At $T=8.765$~K for different currents. The horizontal dash line indicated that the temperature (8.759~K) or current ($200~\mu$A) where the maxima of resistance in high field regime (e.g. $2.6H_1< H < 4.3H_1$ ) are smaller than those in low field regime ($H < 2.5H_1$).}
  \end{center}
\end{figure}

Since pattern A and B have a similar behavior in magnetoresistance, we shall focus on the $R(H)$ curves of pattern A. In Figure \ref{RH}(a), we measured the $R(H)$ curves at several temperatures from 8.794 K to 8.740 K, with a fixed current $I$ = 250 $\mu$A. The reduced temperature $t$ = $T/T_c$ ranged from 0.998 to 0.992. Obviously, the saturation number is temperature dependent\cite{Mkrtchyan.jetp}. We found $N_{SA}$ = 2 at $T$ = 8.794 K but $N_{SA}$ = 3 at $T$ = 8.740 K. We also found a different temperature dependence of the $R(H)$ curves between the low field and high field regime, suggesting varied pinning mechanisms involved. In low field regime, the matching behavior showed a stability in a wide range of temperature. From 8.794 K to 8.740 K, the minima in integer matching fields can always be visible due to the strong pinning by large holes. However, the dips in the high field regime were strongly temperature dependent, indicating high mobility of interstitial vortices. In general, the entrance of interstitial vortices would cause a drastic increase in magnetoresistance due to their weak pinning potentials\cite{Rosseel_prb_1996}. For our samples, at temperature close to $T_c$ (such as $T$ $>$ 8.776 K), the background of resistance raised up as the magnetic field was increased, which was common seen in the periodic pinning arrays\cite{Hoffmann_prb_2000}. In contrast, when the temperature was below a certain value (near 8.759 K), abnormal resistance behavior occurred. The resistance in high field kept in a lower level than the low field regime, which also can be found in rectangular arrays of magnetic dots\cite{Artificially,Temperature,Hysteresis}. The lower temperature, the broader range of the abnormal behavior can be visible. To further study the phenomenon, we examined current dependence of magnetoreistance at a fixed temperature $T$ = 8.765 K in Figure \ref{RH}(b). The $R(H)$ curves were recorded with currents in the range from 50 $\mu$A to 450 $\mu$A, corresponding to the current density ranged from 4.03 to 48.36 kA/cm$^2$ which well below the depairing critical current. The current dependence of the $R(H)$ curves are much like the temperature dependence shown in Figure \ref{RH}(a). The abnormal magnetoresistance behavior occurred at currents below $200$~$\mu$A, and revealed an enhancement of the critical current in high field. Thus, the abnormal magnetoresistance behavior is highly influenced by temperature and current.

The strong temperature and current dependence of magnetoresistance in high field regime indicates the competition between the pinning potentials and driving force\cite{Reichhardt_prb_2001}. The pining potentials of interstitial vortices are dominant by the strong repulsive interactions with multivortices pinned in holes, which caged the interstitial vortices effectively in a certain range of temperatures. 
This 'caging effect' has been predicted by theoretical simulations\cite{Berdiyorov_EPL_2006, Berdiyorov_prb_2006} and found greatly temperature dependent.
In the simulations, they predicted a recovering of normal behavior at temperatures far below $T_c$ ($T/T_c < 0.8$). It is difficult to check this recovering phenomenon in our magnetoresistance measurements due to a requirement of large applied current in low temperatures. A direct measurement of $J_c(H)$ will be necessary to further confirm this effect. In the other hand, the decreasing of current density reduced the driving force, resulting in effective pinning of vortices by the pinning potentials. In both cases, the lowering of mobility of interstitial vortices causes the reduction in magnetoresistance. Therefore, the abnormal behavior is owing to the interstitial vortices caged among the large number of pinned ones. Another possible explanation is due to the enhanced rigidity of vortex lattice in high field, where the lattice cannot easily flow around pinned vortices.\cite{Temperature}

\section{Conclusion}
We investigated the magnetoresistance of superconducting Nb thin films containing rectangular arrays of antidots with different aspect ratios. We found that the reconfiguration transition highly depends on the antidot saturation number $N_S$. The latter is determined by the size of the antidots, their spacing and temperature. After the transition, resistance (critical current) becomes smaller (larger), in contrast to the conventional behavior. We attribute the abnormal behavior to the lowering of vortex mobility in high field regime where interstitial vortices were strongly caged by pinning multivortices. This effect is affected by temperature and current, which consists with recent theoretical predictions.


\end{document}